\documentclass[a4paper,conference,10pt,table]{IEEEtran}
\IEEEoverridecommandlockouts
\usepackage{amsmath,amssymb,amsfonts}
\usepackage[numbers,sort&compress]{natbib}
\usepackage{algorithm}
\usepackage{algorithmic}
\usepackage{graphicx}
\usepackage{textcomp}
\usepackage{comment}
\usepackage{subcaption}
\usepackage{amsmath}       
\usepackage{array}        
\usepackage[table,xcdraw]{xcolor} 
\usepackage{graphicx}      
\usepackage{caption}
\usepackage[table,xcdraw]{xcolor}
\usepackage{booktabs} 
\usepackage[acronyms,nonumberlist,nopostdot,nomain,nogroupskip]{glossaries}
\newcommand{\red}[1]{\textcolor{red}{#1}}

\def\tx{\mathrm{tx}}
\def\rx{\mathrm{rx}}

\makeatletter
\let\oldabs\abs
\def\abs{\@ifstar{\oldabs}{\oldabs*}}

\let\oldnorm\norm
\def\norm{\@ifstar{\oldnorm}{\oldnorm*}}
\makeatother
\renewcommand{\a}{\mathbf{a}}
\renewcommand{\b}{\mathbf{b}}

\renewcommand{\d}{\mathbf{d}}

\newcommand{\h}{\mathbf{h}}

\newcommand{\m}{\mathbf{m}}

\newcommand{\p}{\mathbf{p}}

\newcommand{\s}{\mathbf{s}}

\renewcommand{\u}{\mathbf{u}}
\renewcommand{\v}{\mathbf{v}}
\newcommand{\w}{\mathbf{w}}
\newcommand{\x}{\mathbf{x}}

\newcommand{\0}{\mathbf{0}}
\newcommand{\1}{\mathbf{1}}

\newcommand{\A}{\mathbf{A}}

\newcommand{\E}{\mathbf{E}}

\newcommand{\G}{\mathbf{G}}
\renewcommand{\H}{\mathbf{H}}

\newcommand{\M}{\mathbf{M}}

\newcommand{\V}{\mathbf{V}}
\newcommand{\W}{\mathbf{W}}

\newcommand{\thetab}{\boldsymbol{\theta}}

\newcommand{\defeq}{\stackrel{\triangle}{=}}

\newcommand{\Compl}{\mbox{$\mathbb{C}$}}

\newcommand{\diag}{\mathrm{diag}}

\newcommand{\herm}{\mathrm{H}}

\renewcommand{\Re}{\mathrm{Re}}
\newcommand{\tr}{\mathrm{tr}}
\newcommand{\tran}{\mathrm{T}}

\newcommand{\MD}[1]{{\color{blue}MD: #1}}

\newacronym{3gpp}{3GPP}{3rd Generation Partnership Project}
\newacronym{5g}{5G}{5th generation}
\newacronym{6g}{6G}{6th generation}
\newacronym{adc}{ADC}{Analog to Digital Converter}
\newacronym{an}{AN}{Artificial Noise}
\newacronym{ao}{AO}{Alternating Optimization}
\newacronym{aoa}{AoA}{Angle of Arrival}
\newacronym{aod}{AoD}{Angle of Departure}
\newacronym{b5g}{B5G}{Beyond-5th generation}
\newacronym{bcd}{BCD}{Block Coordinate Decent}
\newacronym{4g}{4G}{4th generation}
\newacronym{aimd}{AIMD}{Additive Increase Multiplicative Decrease}
\newacronym{a2g}{A2G}{air-to-ground}
\newacronym{am}{AM}{Acknowledged Mode}
\newacronym{amc}{AMC}{Adaptive Modulation and Coding}
\newacronym{ap}{AP}{Access Point}
\newacronym{aqm}{AQM}{Active Queue Management}
\newacronym{asr}{ASR}{Average Secrecy Rate}
\newacronym{awgn}{AWGN}{additive white Gaussian noise}
\newacronym{balia}{BALIA}{Balanced Link Adaptation}
\newacronym{bdp}{BDP}{Bandwidth-Delay Product}
\newacronym{bf}{BF}{Beamforming}
\newacronym{cc}{CC}{Congestion Control}
\newacronym{cu}{CU}{Central Unit}
\newacronym{ecdf}{ECDF}{Empirical Cumulative Distribution Function}
\newacronym{cn}{CN}{Core Network}
\newacronym{cqi}{CQI}{Channel Quality Information}
\newacronym{cp}{CP}{Control Plane}
\newacronym{csirs}{CSI-RS}{Channel State Information - Reference Signal}
\newacronym{dc}{DC}{Dual Connectivity}
\newacronym{dce}{DCE}{Direct Code Execution}
\newacronym{dci}{DCI}{Downlink Control Information}
\newacronym{drl}{DRL}{Deep Reinforcement Learning}
\newacronym{dl}{DL}{Downlink}
\newacronym{du}{DU}{Distributed Unit}
\newacronym{dmr}{DMR}{Deadline Miss Ratio}
\newacronym{dmrs}{DMRS}{DeModulation Reference Signal}
\newacronym{e2e}{E2E}{end-to-end}
\newacronym{ecn}{ECN}{Explicit Congestion Notification}
\newacronym{edf}{EDF}{Earliest Deadline First}
\newacronym{enb}{eNB}{evolved Node Base}
\newacronym{epc}{EPC}{Evolved Packet Core}
\newacronym{es}{ES}{Edge Server}
\newacronym{fdma}{FDMA}{Frequency Division Multiple Access}
\newacronym{fdd}{FDD}{Frequency Division Duplexing}
\newacronym[firstplural=Radio Access Technologies (RATs)]{rat}{RAT}{Radio Access Technology}
\newacronym{fs}{FS}{Fast Switching}
\newacronym{ftp}{FTP}{File Transfer Protocol}
\newacronym{gnb}{gNB}{Next Generation Node Base}
\newacronym{harq}{HARQ}{Hybrid Automatic Repeat reQuest}
\newacronym{hetnet}{HetNet}{Heterogeneous Network}
\newacronym{hh}{HH}{Hard Handover}
\newacronym{hol}{HOL}{Head-of-Line}
\newacronym{ia}{IA}{Initial Access}
\newacronym{ieee}{IEEE}{Institute of Electrical and Electronics Engineers}
\newacronym{ilp}{ILP}{Integer Linear Program}
\newacronym{imt}{IMT}{International Mobile Telecommunication}
\newacronym{itu}{ITU}{International Telecommunication Union}
\newacronym{iot}{IoT}{Internet of Things}
\newacronym{isac}{ISAC}{Integrated Sensing and Communication}
\newacronym{kkt}{KKT}{Karush-Kuhn-Tucker}
\newacronym{ldpc}{LDPC}{Low-Density Parity Check}
\newacronym{lc}{LC}{Liquid Crystal}
\newacronym{lcd}{LCD}{Liquid Crystal Display}
\newacronym{los}{LoS}{Line-of-Sight}
\newacronym{lrs}{LRS}{Large Reflecting Surface}
\newacronym{lte}{LTE}{Long Term Evolution}
\newacronym{m2m}{M2M}{Machine to Machine}
\newacronym{mac}{MAC}{Medium Access Control}
\newacronym{mc}{MC}{Multi-Connectivity}
\newacronym{mcs}{MCS}{Modulation and Coding Scheme}
\newacronym{mec}{MEC}{Mobile Edge Cloud}
\newacronym{mems}{MEMS}{micro-electromechanical systems}
\newacronym{mi}{MI}{Mutual Information}
\newacronym{mimo}{MIMO}{multiple-input multiple-output}
\newacronym{miso}{MISO}{multiple-input single-output}
\newacronym{mmwave}{mmWave}{millimeter-wave}
\newacronym{mptcp}{MPTCP}{Multipath TCP}
\newacronym{mr}{MR}{Maximum Rate}
\newacronym{mss}{MSS}{Maximum Segment Size}
\newacronym{mtd}{MTD}{Machine-Type Device}
\newacronym{mtu}{MTU}{Maximum Transmission Unit}
\newacronym{nfv}{NFV}{Network Function Virtualization}
\newacronym{nlos}{NLoS}{Non-Line-of-Sight}
\newacronym{nlosv}{NLOSv}{Vehicle Non-Line-of-Sight}
\newacronym{nr}{NR}{New Radio}
\newacronym{ofdm}{OFDM}{Orthogonal Frequency Division Multiplexing}
\newacronym{pdcch}{PDCCH}{Physical Downlonk Control Channel}
\newacronym{pdcp}{PDCP}{Packet Data Convergence Protocol}
\newacronym{pdsch}{PDSCH}{Physical Downlink Shared Channel}
\newacronym{pdu}{PDU}{Packet Data Unit}
\newacronym{psd}{PSD}{positive semi-definite}
\newacronym{pf}{PF}{Proportional Fair}
\newacronym{pgw}{PGW}{Packet Gateway}
\newacronym{phy}{PHY}{Physical}
\newacronym{pbch}{PBCH}{Physical Broadcast Channel}
\newacronym{pla}{PLA}{Planar Linear Array}
\newacronym{pls}{PLS}{Physical Layer Security}
\newacronym[plural=\gls{mme}s,firstplural=Mobility Management Entities (MMEs)]{mme}{MME}{Mobility Management Entity}
\newacronym{prb}{PRB}{Physical Resource Block}
\newacronym{pss}{PSS}{Primary Synchronization Signal}
\newacronym{pscch}{PSCCH}{Physical Sidelink Control Channel}
\newacronym{pucch}{PUCCH}{Physical Uplink Control Channel}
\newacronym{pusch}{PUSCH}{Physical Uplink Shared Channel}
\newacronym{rach}{RACH}{Random Access Channel}
\newacronym{ran}{RAN}{Radio Access Network}
\newacronym{red}{RED}{Random Early Detection}
\newacronym{rf}{RF}{radio frequency}
\newacronym{rlc}{RLC}{Radio Link Control}
\newacronym{rlf}{RLF}{Radio Link Failure}
\newacronym{rrc}{RRC}{Radio Resource Control}
\newacronym{rrm}{RRM}{Radio Resource Management}
\newacronym{rr}{RR}{Round Robin}
\newacronym{rs}{RS}{Remote Server}
\newacronym{rsrp}{RSRP}{Reference Signal Received Power}
\newacronym{rss}{RSS}{Received Signal Strength}
\newacronym{rtt}{RTT}{Round Trip Time}
\newacronym{rw}{RW}{Receive Window}
\newacronym{rx}{RX}{Receiver}
\newacronym{sa}{SA}{standalone}
\newacronym{sack}{SACK}{Selective Acknowledgment}
\newacronym{sap}{SAP}{Service Access Point}
\newacronym{sc}{SC}{Single Carrier}
\newacronym{sch}{SCH}{Secondary Cell Handover}
\newacronym{scoot}{SCOOT}{Split Cycle Offset Optimization Technique}
\newacronym{sdma}{SDMA}{Spatial Division Multiple Access}
\newacronym{sdr}{SDR}{semi-definite relaxation}
\newacronym{sinr}{SINR}{signal-to-interference-plus-noise ratio}
\newacronym{siso}{SISO}{Single-Input-Single-Output}
\newacronym{sl}{SL}{Sidelink}
\newacronym{slnr}{SLNR}{Signal-to-Leakage-and-Noise-Ratio}
\newacronym{sm}{SM}{Saturation Mode}
\newacronym{snr}{SNR}{Signal-to-Noise-Ratio}
\newacronym{son}{SON}{Self-Organizing Network}
\newacronym{ss}{SS}{Synchronization Signal}
\newacronym{srs}{SRS}{Sounding Reference Signal}
\newacronym{sss}{SSS}{Secondary Synchronization Signal}
\newacronym{tb}{TB}{Transport Block}
\newacronym{tcp}{TCP}{Transmission Control Protocol}
\newacronym{tdd}{TDD}{Time Division Duplexing}
\newacronym{thz}{THz}{Terahertz}
\newacronym{tdma}{TDMA}{Time Division Multiple Access}
\newacronym{tfl}{TfL}{Transport for London}
\newacronym{tm}{TM}{Transparent Mode}
\newacronym{trp}{TRP}{Transmitter Receiver Pair}
\newacronym{tti}{TTI}{Transmission Time Interval}
\newacronym{ttt}{TTT}{Time-to-Trigger}
\newacronym{tx}{TX}{Transmitter}
\newacronym{ue}{UE}{user equipment}
\newacronym{ul}{UL}{Uplink}
\newacronym{uml}{UML}{Unified Modeling Language}
\newacronym{um}{UM}{Unacknowledged Mode}
\newacronym{utc}{UTC}{Urban Traffic Control}
\newacronym{vm}{VM}{Virtual Machine}
\newacronym{rsrq}{RSRQ}{Reference Signal Received Quality}
\newacronym{rssi}{RSSI}{Received Signal Strength Indicator}
\newacronym{rv}{RV}{Random Variable}
\newacronym{crs}{CRS}{Cell Reference Signal}
\newacronym{nsa}{NSA}{Non Stand Alone}
\newacronym{noma}{NOMA}{Non-orthogonal Multiple Access}
\newacronym{mrdc}{MR-DC}{Multi \gls{rat} \gls{dc}}
\newacronym{endc}{EN-DC}{E-UTRAN-\gls{nr} \gls{dc}}
\newacronym{5gc}{5GC}{5G Core}
\newacronym{si}{SI}{Study Item}
\newacronym{iab}{IAB}{Integrated Access and Backhaul}
\newacronym{wf}{WF}{Wired-first}
\newacronym{hqf}{HQF}{Highest-quality-first}
\newacronym{pa}{PA}{Position-aware}
\newacronym{mlr}{MLR}{Maximum-local-rate}
\newacronym{wbf}{WBF}{Wired Bias Function}
\newacronym{mib}{MIB}{Master Information Block}
\newacronym{sca}{SCA}{successive convex approximation}
\newacronym{sib}{SIB}{Secondary Information Block}
\newacronym{rnti}{RNTI}{Radio Network Temporary Identifier}
\newacronym{dft}{DFT}{Discrete Fourier Transform}
\newacronym{kpi}{KPI}{Key Performance Indicator}
\newacronym{ppp}{PPP}{Poisson Point Process}
\newacronym{v2v}{V2V}{Vehicle-to-Vehicle}
\newacronym{wave}{WAVE}{Wireless Access in Vehicular Environments}
\newacronym{udp}{UDP}{User Datagram Protocol}
\newacronym{upa}{UPA}{uniform planar array}
\newacronym{fec}{FEC}{Forward Error Correction}
\newacronym{v2x}{V2X}{Vehicle-To-Everything}
\newacronym{psfch}{PSFCH}{Physical Sidelink Feedback Channel}
\newacronym{pssch}{PSSCH}{Physical Sidelink Shared Channel}
\newacronym{csma}{CSMA}{Carrier Sense Multiple Access}
\newacronym{v2n}{V2N}{Vehicle-to-Network}
\newacronym{wlan}{WLAN}{Wireless Local Area Network}
\newacronym{cav}{CAV}{Connected and Autonomous Vehicle}
\newacronym{v2i}{V2I}{Vehicle-to-Infrastructure}
\newacronym{d2d}{D2D}{Device-to-Device}
\newacronym{c-its}{C-ITS}{Connected Intelligent Transportation System}
\newacronym{fr2}{FR2}{Frequency Range 2}
\newacronym{bs}{BS}{base station}
\newacronym{sdu}{SDU}{Service Data Unit}
\newacronym{csi}{CSI}{Channel State Information}
\newacronym{scs}{SCS}{Subcarrier Spacing}
\newacronym{sop}{SOP}{Secrecy Outage Probability}
\newacronym{sumo}{SUMO}{Simulation of Urban MObility}
\newacronym{prr}{PRR}{Packet Reception Ratio}
\newacronym{edca}{EDCA}{Enhanced Distribution Channel Access}
\newacronym{sdap}{SDAP}{Service Data Adaptation Protocol}
\newacronym{sdp}{SDP}{Semi-definite Programming}
\newacronym{scm}{SCM}{Spatial Channel Model}
\newacronym{vr}{VR}{Virtual Reality}
\newacronym{qos}{QoS}{Quality of Service}
\newacronym{uav}{UAV}{unmanned Aerial Vehicle}
\newacronym{bap}{BAP}{Backhaul Adaptation Protocol}
\newacronym{ns3}{ns-3}{Network Simulator 3}
\newacronym{rl}{RL}{Reinforcement Learning}
\newacronym{ris}{RIS}{reconfigurable intelligent surface}
\newacronym{iris}{IRIS}{Illegal Reconfigurable Intelligent Surface}
\newacronym{ula}{ULA}{uniform linear array}
\newacronym{pin}{PIN}{positive-intrinsic-negative}
\newacronym{mrt}{MRT}{maximum ratio transmission}
\usepackage{url}
\usepackage[T1]{fontenc}
\usepackage[utf8]{inputenc}
\usepackage[inline]{enumitem} 

\glsdisablehyper
\usepackage{flushend}
\usepackage[letterpaper,top=0.7in,bottom=0.98in,left=0.625in,right=0.625in]{geometry}

\def\BibTeX{{\rm B\kern-.05em{\sc i\kern-.025em b}\kern-.08em
    T\kern-.1667em\lower.7ex\hbox{E}\kern-.125emX}}
    
\begin{document}

\title{Temperature-Resilient LC-RIS Phase-Shift Design for Multi-user Downlink Communications
}

\author{
    \IEEEauthorblockN{
        Nairy Moghadas Gholian\IEEEauthorrefmark{1},
        Mohamadreza Delbari\IEEEauthorrefmark{1},
        Vahid Jamali\IEEEauthorrefmark{1},
        and Arash Asadi\IEEEauthorrefmark{2}
    }
    \IEEEauthorblockA{
        \IEEEauthorrefmark{1}Technical University of Darmstadt (TUDa), Darmstadt, Germany \\
        ngholian@seemoo.tu-darmstadt.de, \{mohamadreza.delbari, vahid.jamali\}@tu-darmstadt.de \\
        \IEEEauthorrefmark{2}Technical University of Delft, Delft, Netherlands \\
        a.asadi@tudelft.nl
    }
    \thanks{Moghadas Gholian and Asadi’s work was funded by the German Research Foundation (DFG) through the project HyRIS (Grant no. 455077022). Delbari and Jamali’s work was supported in part by the Deutsche Forschungsgemeinschaft (DFG, German Research Foundation) within the Collaborative Research Center MAKI (SFB 1053, Project-ID 210487104) and in part by the LOEWE initiative (Hesse, Germany) within the emergenCITY Centre under Grant LOEWE/1/12/519/03/05.001(0016)/72.}
    }

\maketitle

\begin{abstract}
The reflecting antenna elements in most \glspl{ris} use semiconductor-based (e.g., \gls{pin} diodes and varactors) phase shifters. Although effective, a drawback of this technology is the high power consumption and cost, which become particularly prohibitive in \gls{mmwave}/sub-Terahertz range. With the advances in \glspl{lc} in microwave engineering, we have observed a new trend in using \gls{lc} for realizing phase shifter networks of \glspl{ris}. \gls{lc}-\glspl{ris} are expected to significantly reduce the fabrication costs and power consumption. However, the nematic \gls{lc} molecules are sensitive to temperature variations. Therefore, implementing \gls{lc}-\gls{ris} in geographical  
regions with varying temperatures requires temperature-resilient designs. The mentioned temperature variation issue becomes more significant at higher temperatures as the phase shifter range reduces in warmer conditions, whereas it expands in cooler ones. In this paper, we study the impact of temperature on the operation of \gls{lc}-\glspl{ris} and develop a temperature-resilient phase shift design. Specifically, we formulate a max-min signal-to-interference-plus-noise ratio optimization for a multi-user downlink \gls{mmwave} network that accounts for the impact of temperature in the \gls{lc}-\gls{ris} phase shifts. The simulation results demonstrate a significant improvement for the considered set of parameters when using our algorithm compared to the baseline approach, which neglects the temperature effects.
\end{abstract}
\glsresetall
\section{Introduction}
With the advent of the sixth generation (6G) wireless communication, \gls{ris} technology has been extensively investigated. These surfaces, typically consist of numerous reconfigurable elements that dynamically adjust the phase of reflected signals, enabling adaptive control of wireless propagation environments \cite{Wu2019}. \glspl{ris} are envisioned to mitigate communication challenges such as blockage, limited coverage, and security threats while simultaneously enhancing energy efficiency \cite{Huang2019, Yu2020,yu2021smart, Gholian2023}. 
Despite their potential, \gls{ris} implementations employing semiconductor-based phase shifters, such as \gls{pin} diodes and varactors, present significant drawbacks, particularly in the \gls{mmwave} and sub-Terahertz (THz) ranges. These semiconductor-based solutions typically entail high power consumption and elevated manufacturing costs, hindering large-scale deployment \cite{Dai2020, Gros2021}.



\gls{lc} technology has been proposed as a promising alternative for the implementation of \gls{ris} phase shifter networks due to its cost-effectiveness, lower power consumption, and capability to provide continuous phase shift adjustments~\cite{Guirado2022, Tesmer2021}. This technology was initially recognized for its extensive use in \glspl{lcd}. Beyond its well-known application in displays, \gls{lc} offers several non-display use cases. Notably, \gls{lc} is a mature, cost-effective technology backed by a robust manufacturing ecosystem, making it an attractive solution for tunable phase shifters compared to silicon-based technologies, such as radio frequency (RF) switches and \gls{pin} diode \glspl{ris} \cite{Zhang2022,Guirado2022,jimenez2023reconfigurable,delbari2024fast}. Additionally, it provides continuous phase shift output at each cell, allowing for finer tuning than the discrete phase shifts typically offered by silicon-based solutions. 
However, one critical but often overlooked challenge in \gls{lc}-\glspl{ris} is their sensitivity to ambient temperature variations. More specifically, nematic \gls{lc} molecules experience significant changes in their electromagnetic properties with temperature fluctuations, directly affecting their phase-shifting capabilities. This thermal sensitivity becomes particularly problematic for \gls{lc}-\glspl{ris} that are designed for a fixed temperature. When the temperature rises, the operational phase shift range of the \gls{lc}-\gls{ris} elements is reduced, limiting the system's overall performance and flexibility.

Despite these clear, practical implications, the existing literature lacks comprehensive studies addressing temperature-induced performance degradation in \gls{lc}-\gls{ris} systems. Recent works have explored various \gls{lc}-\gls{ris} characteristics, such as hardware design challenges \cite{jimenez2023reconfigurable}, slow transition times \cite{delbari2024fast}, and secrecy issues under thermal effects \cite{Delbari2024temperature}. Nonetheless, to the best of our knowledge, no prior work has investigated the impact of temperature variation on multi-user downlink \gls{lc}-\gls{ris}-assisted networks.
To fill this gap, in this paper, we propose a novel temperature-resilient \gls{lc}-\gls{ris} design framework specifically aimed at maximizing the minimum \gls{sinr} in a multi-user \gls{miso} \gls{mmwave} network. In the following, we summarize our contributions.

\begin{itemize}
    \item We propose a novel multi-user downlink system model that incorporates the impact of ambient temperature variation on the performance and fairness of \gls{lc}-\gls{ris} in a \gls{miso} network. Specifically, we analyze the effect of varying temperature on the phase shifter range in a system with multiple \glspl{ue} within a predefined region around the \gls{lc}-\gls{ris}.

    \item Next, we design the phase shifter of an \gls{lc}-\gls{ris}-assisted network in a downlink multi-user \gls{miso} \gls{mmwave} network with attenuated \gls{los} link. In our model, as the temperature fluctuates, the phase shifter range of the \gls{lc}-\gls{ris} becomes confined to only a fraction of the full $[0, 2\pi]$ range. We propose an optimization framework to handle this constraint and ensure system fairness across all users.
    
    \item {Lastly, we evaluate our proposed scheme under elevated temperature conditions, comparing it against three different schemes. Our evaluations demonstrate that, even with temperature fluctuations, our proposed scheme achieves a max-min \gls{sinr} value that is significantly improved compared to the benchmark schemes. 
}
\end{itemize}
The only existing work that has studied the impact of temperature in \gls{lc}-\gls{ris} phase shift design is \cite{Delbari2024temperature}. While \cite{Delbari2024temperature} focuses on optimizing the secrecy rate in a single-user scenario, our objective is to maximize the minimum \gls{sinr} in a multi-user downlink setup. Additionally, \cite{Delbari2024temperature} employs a \gls{sdr} approach to address the phase shifter problem at the \gls{lc}-\gls{ris}. Although \gls{sdr} is effective for non-convex problems, its computational complexity ($\mathcal{O}(N^3)$, with $N$ denoting the number of elements) limits the applicability of the algorithm developed in \cite{Delbari2024temperature} to \gls{lc}-\glspl{ris} with only a few hundred elements. In contrast, our proposed method uses a \gls{sca} approach with a computational complexity of $\mathcal{O} (N)$ to optimize the phase shifters, enabling us to analyze systems with a much larger number of \gls{ris} elements and demonstrating the scalability and effectiveness of our proposed approach.

\textit{Notation:} Matrices and vectors are denoted in uppercase and lowercase bold font, respectively. $(\cdot)^{\tran}$, $(\cdot)^{*}$, $(\cdot)^{\herm}$, $\tr(\cdot)$, and $\text{Rank}(\cdot)$ stand for transposition, complex conjugation, Hermitian transposition, trace, and rank of a square matrix, respectively. Additionally, $|\cdot |$ denotes the absolute value of a complex number. Moreover, $\mathsf{i}=\sqrt{-1}$ is the imaginary number, and $\mathrm{diag}(\x)$ is a square matrix whose diagonal is equal to $\x$ and all other elements are zero. $\Re{\{\cdot\}}$ stands for the real part of a complex number and $[\a]_m$ represents the $m$-th entry of the vector $\a$. Finally, $\mathcal{O}$ stands for the big-O notation and $\{\cdot\} \succeq 0$ indicates a \gls{psd} matrix.
\section{System and Channel Models}
\subsection{System model}
\begin{figure}[t]
    \centering
    \includegraphics[width=1\columnwidth]{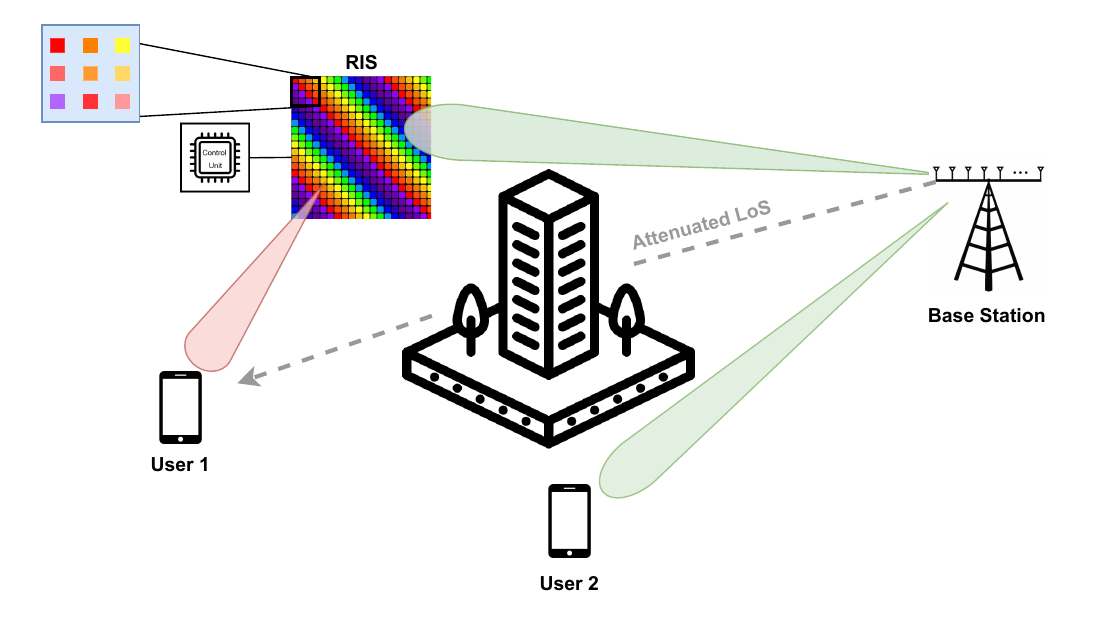}
    \caption{An \gls{lc}-\gls{ris} assisted \gls{mmwave} wireless network, where the \gls{los} link between \gls{bs} and \gls{ue} is blocked.}
    \label{fig:min_sinr_vs_users}
    \vspace{-3ex}
\end{figure}

We consider a multi-user \gls{miso} downlink system operating at \gls{mmwave} frequencies, assisted by an \gls{lc}-\gls{ris} as illustrated in Fig.~\ref{fig:min_sinr_vs_users}. The network consists of a \gls{bs} equipped with $M$-element \gls{ula} spread across the x-axis, serving $K$ single-antenna \glspl{ue} through an \gls{lc}-\gls{ris}. The \gls{lc}-\gls{ris} deployed as a \gls{upa} comprises $N = N_x \times N_z$ antenna elements, where $N_x$ and $N_z$ denote the numbers of reflecting elements arranged along the horizontal (x-axis) and vertical (z-axis) dimensions, respectively. Both the \gls{bs} and \gls{ris} antenna arrays maintain half-wavelength spacing $d_c = \lambda/2$, where $\lambda$ denotes the carrier wavelength.
The \gls{bs} simultaneously transmits independent data symbols $s_k\, \in \Compl$ for each \gls{ue} $k$, assuming normalized power such that $\mathbb{E}\{{|s_k|}^2 \} = 1,\,\forall k$. To effectively manage interference and maximize system performance, the \gls{bs} employs a precoding matrix $\W = [ \w_1, \w_2, \cdots, \w_K] \in \Compl ^{M \times K}$ where vector $\w_k \in \Compl ^{M \times 1}$ is designed for \gls{ue} $k$.
The received signal at the $k$-th \gls{ue} is therefore expressed as
\begin{multline}\label{eq:yk1}
    y_k = (\h_k^\herm \mathbf{\Theta} \G + \h_{\mathrm{d},k}^{\herm})\w_k s_k  \\
    + \sum_{j\neq k} (\h_k^\herm \mathbf{\Theta} \G + \h_{\mathrm{d},k}^{\herm}) \w_j s_j + n_k,
\end{multline}
where $\G\in\Compl^{N \times M}$, $\h_k\in\Compl^{N \times 1}$, and $\h_{\mathrm{d},k}\in \Compl^{M \times 1}$ represent the \gls{bs}-to-\gls{ris} channel, \gls{ris}-to-\gls{ue} $k$ channel, and the direct \gls{bs}-to-\gls{ue} $k$ channel,  respectively.
Additionally, $n_k$ is the \gls{awgn} at the $k$th \gls{ue}, which follows the Gaussian random distribution $n_{k} \sim \mathcal {CN}(0, \sigma ^{2}_{k})$ where $ \sigma ^{2}_{k}$ is the noise variance and $\mathbf{\Theta} = {\rm{diag}}\left( {{\beta _1}{e^{\mathsf{i}{\theta _1}}},...,{\beta _N}{e^{\mathsf{i}{\theta_N}}}} \right) \in {\mathbb{C}^{N \times N}}$ where $\beta_n$ and $\theta_n,\,\forall n\in\{1, \cdots, N\}$ are amplitude and phase shifter of $n$th cell in LC-RIS, respectively. For passive \gls{ris} assumption and following practical \gls{lc}-\gls{ris} designs, we assume unit amplitude reflection from each \gls{lc}-\gls{ris} element meaning that $|\beta_n| = 1$ \cite{delbari2024fast}. Moreover, for further simplicity, we define the overall phase shifter vector as $\boldsymbol {\theta }\defeq [\theta _{1}, \cdots, \theta_{N}]$. 

\subsection{Channel model}
We adopt a Rician fading model due to the presence of both \gls{los} and \gls{nlos} paths at \gls{mmwave} bands. Under this model, any link between transmitter with $N_{\tx}$ antennas and a receiver with $N_{\rx}$ antennas is modeled as
\begin{equation}\label {eq:rician1}
\H \triangleq \sqrt {\frac {K_{\mathrm{t,r}}}{1+K_{\mathrm{t,r}}}} \, \H ^{\mathrm{LoS}} + \sqrt {\frac {1}{1+K_{\mathrm{t,r} }}}\,\H^{ \mathrm {NLoS}} \in { \mathbb {C}} ^{N_{\rx}\times N_\tx},
\end{equation}
where $K_{\mathrm{t,r}}$ is the Rician factor, representing the ratio between the power of \gls{los} and \gls{nlos} components. $\H^\mathrm{LoS}$ and $\H^\mathrm{NLoS}$ denote the deterministic \gls{los} channel component and the random \gls{nlos} channel component, respectively.
For the channel $\G$, $\h_k$, and $\h_{\mathrm{d},k}$, we independently assign corresponding Rician factors, denoted as $K_{\mathrm{R}}$, $K_{\mathrm{R},k}$, and $K_{\mathrm{R,d},k}$, respectively.

\section{Temperature-resilient LC-RIS Phase-Shift Design}
In this section, we begin with an overview of the principle of the \gls{lc} phase shifter. Subsequently, we model the effect of temperature variations on the phase shift of each \gls{lc}-\gls{ris} cell \cite{Delbari2024temperature}. We then formulate an optimization problem aimed at maximizing the minimum \gls{sinr} user while incorporating the constraints imposed by temperature fluctuations. Specifically, we account for the temperature-induced limitation on the maximum achievable phase shift by including it as a constraint in the problem formulation.
\subsection{An overview of the LC phase shifter principle}
\begin{figure}[ht]
    \centering
    \begin{subfigure}[c]{0.48\linewidth}
        \centering       \includegraphics[width=1\linewidth, trim=0cm 0cm 0cm 0, clip]{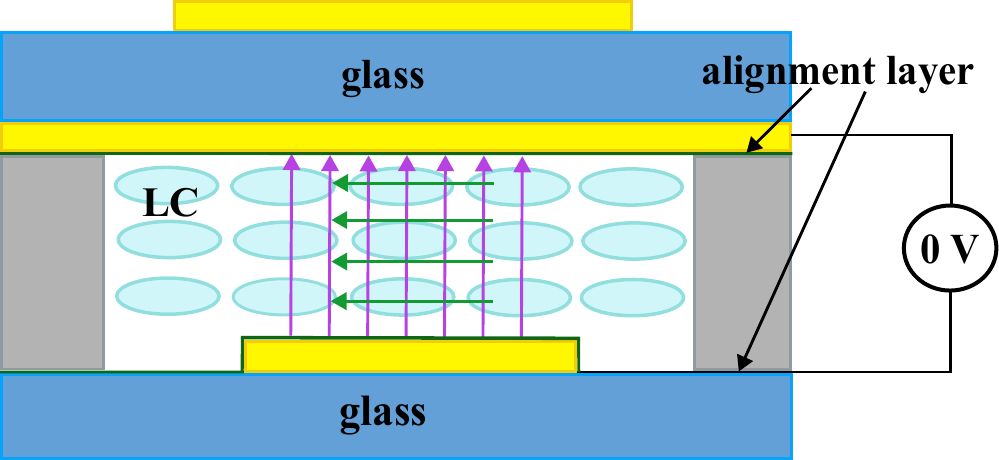}
        \caption{}
        \label{fig:zero_voltage}
    \end{subfigure}
    \hfill
    \begin{subfigure}[c]{0.48\linewidth}
        \centering
        \includegraphics[width=1\linewidth, trim=0cm 0cm 0cm 0, clip]{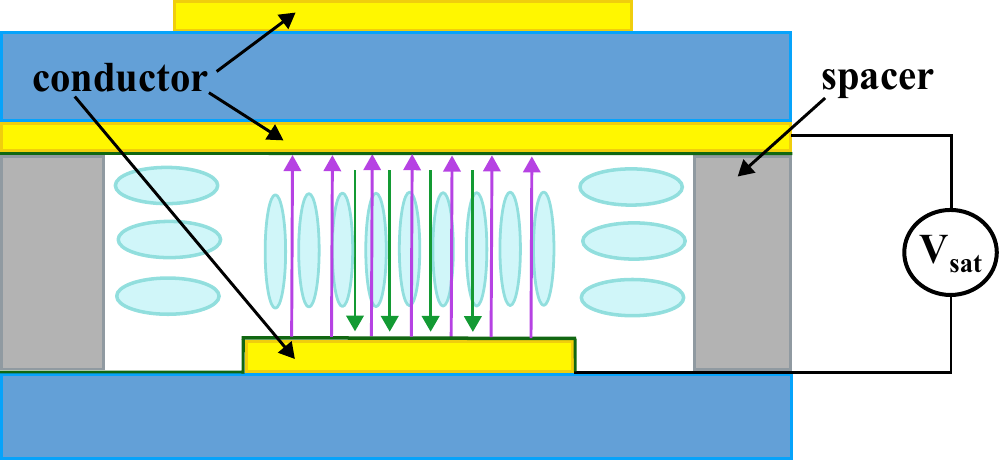}
        \caption{}
        \label{fig:bias_voltage}
    \end{subfigure}
    \caption{\footnotesize A single \gls{lc}-\gls{ris} antenna element shown under two conditions: (a) unbiased, and (b) fully biased, aligning the \gls{lc} molecules for maximum permittivity \cite{Neuder2024architecture}.}
    \label{fig:combined_voltage_conditions}
\end{figure}

\gls{lc}-\gls{ris} consists of numerous rod-shaped nematic \gls{lc} molecules \cite{Tesmer2021}. When an electromagnetic signal impinges on these molecules, each one reflects the signal with a distinct phase. By applying an external bias voltage, the orientation of the molecules, and consequently the resulting phase shift, can be precisely controlled. This tunable property enables the realization of effective phase shifters within \gls{lc}-\gls{ris} structures \cite{Tesmer2021}. As illustrated in Fig.~\ref{fig:combined_voltage_conditions}, the architecture of a single \gls{lc}-\gls{ris} antenna element consists of a layer of nematic \gls{lc}-mixture placed between two dielectric layers \cite{Neuder2024architecture}. The green arrow indicates the director, which represents the average alignment direction of the \gls{lc} molecules, and the purple arrow shows the \gls{rf} electrical field between the conductors shown in yellow. Fig.~\ref{fig:zero_voltage} depicts the orientation of each molecule in the absence of a bias voltage. In contrast, Fig.~\ref{fig:bias_voltage} demonstrates the molecular alignment when a sufficiently large bias voltage is applied, causing the director of most \gls{lc} molecules to align parallel to the field and achieve maximum permittivity.

\subsection{Impact of temperature on LC-RIS unit-cells}

With the temperature alteration, the generated phase shift of each \gls{lc} cell changes. Typically, the \gls{lc}-\gls{ris} is designed such that each \gls{lc} cell can provide any phase in the range of [0, $2\pi]$ at a reference temperature, denoted by $T_\mathrm{r}$. 
However, nematic \gls{lc} molecules exhibit significant sensitivity to temperature, directly affecting their birefringence and, consequently, the achievable phase shift. To address this effect, we start by explaining the temperature functions of \gls{lc} physical parameters and then analyze their impact on \gls{lc} phase shift.
Empirically, the extraordinary and ordinary refractive indices, denoted as $n_{\mathrm{e}}(T)$ and $n_{\mathrm{o}}(T)$, respectively, vary according to the four-parameter model \cite{Wang2005}
\begin{align}
n_{\mathrm{e}}(T) &\approx A - BT +\frac{2{\Delta n}_0}{3} {\left (1- \frac{T}{T_c} \right )} ^{\alpha}, \label{eq:neno 1}\\
n_{\mathrm{o}}(T) &\approx A - BT - \frac{{\Delta n}_0}{3} {\left (1- \frac{T}{T_c} \right)} ^{\alpha}\label{eq:neno 2},
\end{align}
where $A$ and $B$ are the fitting parameters, $T_c$ is the clearing temperature, $\alpha$ is the molecule structure exponent ranging from $0.20$ to $0.25$, and $T$ is the given temperature \cite{Li2004}. Subtracting \eqref{eq:neno 2} from \eqref{eq:neno 1} yields the temperature-dependent birefringence, which is expressed as
\begin{align}\label{eq:nenodelta}
\Delta n (T) = n_e(T) - n_o(T) = {\Delta n}_0 {\left (1- \frac{T}{T_\mathrm{c}} \right)} ^{\alpha},
\end{align}
where ${\Delta n}_0$ stands for the birefringence in the reference temperature. Furthermore, each \gls{lc}-\gls{ris} element consists of an \gls{lc} cell with a cell gap of $d$, whose maximum achievable phase shift $\theta_{\max}(T)$ at a given temperature $T$ is expressed by \cite{Wang2005}
\begin{equation}\label{eq:phasemax1}
\theta_{\max} (T) = \frac{2 \pi d}{\lambda} \Delta n (T).
\end{equation}
For practical design purposes, the \gls{lc} parameters are selected such that at the reference temperature denoted by $T_{\mathrm{r}}$, the full phase shift range of $[0,2\pi]$ is achievable \citep[p. 103]{Wang2005}. Therefore,
\begin{equation}
\label{eq: theta max}
    \theta_{\max} (T_{\mathrm{r}}) = 2 \pi \overset{(a)}{\Rightarrow} \frac{2\pi d}{\lambda} \Delta n (T_{\mathrm{r}}) = 2 \pi,
\end{equation}
where $(a)$ is obtained from substituting $T_{\mathrm{r}}$ in \eqref{eq:phasemax1}. Thus by combining \eqref{eq:nenodelta} and \eqref{eq: theta max}, the following expression for $\Delta n_0$ can be obtained:
\begin{equation}\label{eq:deltan0}
\Delta n_0 = \frac{\lambda}{d} {\left ( 1 - \frac{T_{\mathrm{r}}}{T_{\mathrm{c}}} \right )}^{-\alpha}.
\end{equation}
By substituting \eqref{eq:deltan0} into \eqref{eq:nenodelta} and then inserting the resulting expression into \eqref{eq:phasemax1}, the temperature-dependent maximum phase shift can be compactly expressed as \cite{Delbari2024temperature}
\begin{equation}
\label{eq: Tmax final}
    \theta_{\max}(T)=2\pi \left ( \frac{T_{\mathrm{c}}-T}{T_{\mathrm{c}}-T_{\mathrm{r}}} \right )^\alpha.
\end{equation}
This function represents the relationship between the maximum achievable phase shifter and a given temperature $T$. Practically, an increase in ambient temperature ($T > T_{\mathrm{r}}$) restricts the phase shift range below $2 \pi$, i.e., $\theta_{\max}(T)<2\pi$, and this significantly affects the \gls{lc}-\gls{ris}'s performance. Ignoring these temperature-induced constraints in \gls{ris} configuration and optimization can severely degrade system reliability, particularly when temperature increases beyond $T_\mathrm{r}$.
\subsection{Problem formulation}
Our objective is to optimize the system performance by maximizing the minimum \gls{sinr} across all \glspl{ue} while accounting for the phase shift limitations imposed by temperature on the \gls{lc}-\gls{ris}. We incorporate both the transmit power constraint at the \gls{bs} and temperature-constrained phase shift range at the \gls{lc}-\gls{ris}. To address this, we first rewrite the receive signal expression in \eqref{eq:yk1} as
\begin{multline}\label{eq:yk2}
    y_k = (\v^\herm \H_k + \h_{\mathrm{d},k}^{\herm}) \w_k s_k \\
    + \sum_{j \neq k} (\v^\herm \H_k + \h_{\mathrm{d},k}^{\herm}) \w_j s_j + n_k,
\end{multline}
where $\v = [v_1, \cdots, v_n]^{\herm}\in\Compl^{N\times 1}$ in which $v_n = e^{\mathsf{i} \theta_n}, \; \forall n$ and $\H_k = \mathrm{diag}(\h_k^\herm)\G\in\Compl^{N\times M}$ \cite{Wu2019}. Using \eqref{eq:yk2}, we can model the \gls{sinr} expression for the $k$-th \gls{ue} as follows
\begin{align}\label{eq:sinr}
    \mathrm{SINR}_k = \frac{ |(\v^{\herm} \H_k  + \h_{\mathrm{d},k}^{\herm})\w_k|^2}{\sum_{j\neq k} |( \v^{\herm} \H_k + \h_{\mathrm{d},k}^{\herm})\w_j|^2 + \sigma_k^2} .
\end{align}
We now formulate the optimization problem as
\begin{subequations}\label{eq:maxsnr}
\begin{align}
\mathcal{P}1: \quad & \max_{\W, \boldsymbol{\theta}}  \,\, \min_{k} \,\,  \text{\gls{sinr}}_k  \label{eq:maxsnra} \\ 
\text{s.t.} \quad 
&\text{C1: } \sum_{k=1}^{K} \| \w_k \|^2  \!\leq \!P ,\,\ \forall k, \label{eq:maxsnrb} \\
&\text{C2: } 0 \leq [\boldsymbol{\theta}]_n \leq \theta_\mathrm{max}(T) < 2\pi,\,\ \forall n,  \label{eq:maxsnrc} 
\end{align}  
\end{subequations}
where $P$ stands for the total transmit power budget. In problem $\mathcal{P}1$, we assume that an increase in temperature ($T>T_{\mathrm{r}}$) has caused a reduction in $\theta_{\text{max}}$ to a value below $2\pi$, as quantified by \eqref{eq: Tmax final}. Furthermore, the objective function in $\mathcal{P}1$ is non-convex, primarily due to the \gls{sinr} expression, which involves a ratio of quadratic functions. This type of mathematical expression is inherently non-convex as it does not have a shape that can be optimized using standard convex optimization methods; therefore, it is challenging to be solved. Furthermore, the \gls{sinr} expression depends on both the precoder matrix $\W$ and the phase shift vector $\thetab$, which complicates its optimization even further. In contrast, C1 and C2 are convex in $\W$ and $\thetab$, respectively \cite{boyd2004convex}. To address nonconvexity, we adopt an \gls{ao} technique to solve $\mathcal{P}1$, where the phase shift vector $\boldsymbol{\theta}$ and the precoder matrix $\W$ are optimized in an alternative manner. Specifically, we decompose $\mathcal{P}1$ into two sub-problems. In the first, we optimize $\thetab$ while keeping $\W$ fixed, and in the second, we optimize $\W$ while $\thetab$ is held constant. In the following, we detail each sub-problem and the corresponding optimization strategy.

\textbf{Optimizing the \gls{bs} Precoder:} 
In this stage, we concentrate solely on refining the \gls{bs}'s precoder matrix $\W$ while keeping the \gls{lc}-\gls{ris} phase shifter settings constant. This reduces our task to a well-studied problem in the literature \cite{Wei2024}. To address the non-convexity, we employ an \gls{sdr} approach coupled with the bisection method, and we subsequently solve the resulting problem using the CVX solver \cite{gb08,cvx}.

\textbf{Optimizing \gls{lc}-\gls{ris} Phase shifter:}
 Here, we assume the precoder matrix $\W$ has already been optimized. We now focus on solving problem $\mathcal{P}1$ with the \gls{ris} phase shifter vector $\boldsymbol{\theta}$ as the only variable. Thus, by defining the slack variable $\kappa \geq 0$, the optimization problem is reduced to
\begin{subequations}\label{eq:aophase}
\begin{align}
\mathcal{P}2: \quad & \max_{\boldsymbol{\theta}} \quad\kappa \label{eq:aophasea} \\
\text{s.t.} \quad &\text{C2 and }\text{C3: } \text{SINR}_k \geq \kappa ,\quad \forall k. \label{eq:aophaseb}
\end{align} 
\end{subequations}
The constraint in C3 is non-convex due to the exponential dependence on the phase shifts $[\boldsymbol{\theta}]_n$. Consequently, conventional convex optimization techniques are not directly applicable. Therefore, to address this, we reformulate C3 by applying a first-order Taylor expansion to linearize this constraint. We start with rewriting C3 as
\begin{align}\label{eq:sinr_b}
   { |(\v^{\herm} \H_k  + \h_{\mathrm{d},k}^{\herm})\w_k|^2}  \geq \kappa \left( {\sum_{j\neq k} |( \v^{\herm} \H_k + \h_{\mathrm{d},k}^{\herm})\w_j|^2 + \sigma_k^2} \right).
\end{align}
To isolate the components of the expression that involve $\theta_n$ from those that do not, we introduce the following definitions 
\begin{align}\label{eq:sinr_b_taylor}
    |s_k|^2 -  \kappa \left( \sum_{j\neq k} |s_{k,j}|^2 + \sigma_k^2 \right) \geq 0.
\end{align}
where $s_k$ and $s_{k,j}$ denote the desired signal and interference terms at \gls{ue} $k$, defined as follows
\begingroup
\begin{align} 
    s_k &= \sum_{n=1}^{N} e^{\mathsf{i} \theta_n} a_n^{(k)}  + u_k,  \label{eq:sk1}\\
    s_{k,j} &= \sum_{n=1}^{N} e^{\mathsf{i} \theta_n} a_n^{(k,j)}  + u_{k,j}, \label{eq:skj1}
\end{align}
\endgroup
where $a_n^{(k)} = {[\H_k \w_k]}_n$, $u_k = \h_{\mathrm{d},k}^{\herm} \w_k$,  $a_n^{(k,j)} = {[\H_k \w_j]}_n$, and $u_{k,j} = \h_{\mathrm{d},k}^{\herm} \w_j$. Now, we apply the first-order  Taylor expansion to $e^{\mathsf{i} \theta_n}$ in order to linearize $s_k$ and $s_{k,j}$ around $\theta_n^{(t-1)}$ at iteration $t$, i.e.,
\begin{align} \label{eq:approxim}
  e^{\mathsf{i}\theta_n^{(t)}}
  &\approx e^{\mathsf{i}\theta_n^{(t-1)}} 
      + \mathsf{i}\,e^{\mathsf{i}\theta_n^{(t-1)}}\,\delta\theta_n^{(t)},
\end{align}
where $\delta {\theta_n}^{(t)} = \theta_n^{(t)} - \theta_n^{(t-1)}$.
By inserting~\eqref{eq:approxim} in ~\eqref{eq:sk1} and~\eqref{eq:skj1}, we define $s_k^{(t)}$ and $s_{k,j}^{(t)}$ at iteration $t$, respectively as follows
\begin{align}\label{eq:sk}
    &\!\!\!\!s_k^{(t)}  \!\!\approx \!\!\sum_{n=1}^N \!\left ( e^{\mathsf{i}\theta_n^{(t-1)}}\!\!\!\! + \mathsf{i} e^{\mathsf{i}\theta_n^{(t-1)}} \!\!\delta\theta_n^{(t)}\!\right)\!  a_n^{(k)}\!\! + u_k \!=\! s_k ^{(t-1)} \!\!+\!\delta s_k,\\
    \label{eq:skskj}
    &\!\!\!\!\!\!s_{k,j}^{(t)}\! \approx\!\! \!\sum_{n=1}^N\! \!\left (\! e^{\mathsf{i}\theta_n^{(t-1)}}\!\!\!\! + \!\mathsf{i} e^{\mathsf{i}\theta_n^{(t-1)}} \!\!\delta\theta_n^{(t)}\!\right) \! a_n^{(k,j)}\!\! + \!u_{k,j}\! =\! s_{k,j} ^{(t-1)} \!\!+\!\delta s_{k,j},
\end{align}
where $\delta s_k = \sum_{n=1}^{N} \mathsf{i} e^{\mathsf{i}\theta_n^{(t-1)}} \delta\theta_n^{(t)} a_n^{(k)}$ and $s_k ^{(t-1)} = \sum_{n=1}^{N} e^{\mathsf{i}\theta_n^{(t-1)}} a_n^{(k)} + u_k$ and we can define $\delta s_{k,j}$ and $s_{k,j} ^{(t-1)}$ in a similar way.
Now, using \eqref{eq:sk} and \eqref{eq:skskj} we can expand $|s_k^{(t)}|^2$ and $|s_{k,j}^{(t)}|^2$, respectively as 
\begin{align}\label{eq:taylorsk2}
    |s_k^{(t)}|^2 & \approx {|s_k^{(t-1)}| }^{2}  + {|\delta s_k|}^{2}  +  2 \Re \left\{  s_k^{(t-1) *}  \delta s_k \right\}, \\
\label{eq:taylorskj2}
    |s_{k,j}^{(t)}|^2 & \approx {|s_{k,j}^{(t-1)}| }^{2}  + {|\delta s_{k,j}|}^{2}  +  2 \Re \left\{  s_{k,j}^{(t-1) *}  \delta s_{k,j} \right\},
\end{align}
where we neglect the second-order terms ${|\delta s_k|}^{2}$ and ${|\delta s_{k,j}|}^{2}$ which are negligible for $\delta \theta_n \ll 1 $. Now, By inserting \eqref{eq:taylorsk2} and \eqref{eq:taylorskj2} into \eqref{eq:sinr_b_taylor}, we can express the first-order Taylor series approximation of \eqref{eq:sinr_b_taylor} around $\theta_n^{(t-1)}$ at iteration $t$ as follows and name it as $\widehat{\text{C3}}$
\begin{align}
    &\widehat{\text{C3}}: |s_k^{(t-1)}|^2 - \kappa \left (\sum_{j \neq k} |s_{k,j}^{(t-1)}|^2 - \sigma_k^2 \right ) + \sum_{n=1}^{N} c_n^{(k,j)}   \geq 0,
\end{align}
where each coefficient $c_n^{(k,j)}$ is derived by isolating the linearized contributions of $\delta \theta_n$. Therefore, we have
\begin{align}
    c_n^{(k,j)} = 2 \Bigg[ & \, \Re \left\{  s_{k}^{(t-1)*} \mathsf{i} e^{\mathsf{i}\theta_n^{(t-1)}} a_n^{(k)} \delta \theta_n \right\} \nonumber \\
    & - \kappa \Re \left\{  s_{k,j}^{(t-1)*} \mathsf{i} e^{\mathsf{i}\theta_n^{(t-1)}} a_n^{(k,j)} \delta \theta_n \right\} \Bigg] \cdot
\end{align}
The first-order Taylor expansion of C3 is thus reformulated as a linear inequality, and the updated optimization problem becomes
\begin{subequations}\label{eq:aophase2}
\begin{align}
\mathcal{P}3: \quad & \max_{\thetab} \quad \kappa \label{eq:aophasea2}\\
\text{s.t.}\quad & \text{C2 and $\widehat{\text{C3}}$}, \nonumber
\end{align} 
\end{subequations}
where now both C2 and $\widehat{\text{C3}}$ are linear in $\theta_n,\,\forall n$.
The complete steps of the proposed algorithm are provided in Algorithm \ref{Alg1}.

\begin{footnotesize}
\begin{algorithm}[H]
\caption{Proposed AO algorithm for solving problem $\mathcal{P}1$}
\label{Alg1}
\begin{algorithmic}[1]
\STATE \textbf{Input:} Set iteration counter $i = 0$, initialize phase shifter $\boldsymbol{\theta}^{(i)}$, maximum number of iterations $I_{\max}$, and tolerance $\epsilon$ \\
\STATE \textbf{Output:} Optimal precoder $\W$ and phase shifter $\boldsymbol{\theta}$
\REPEAT
    \STATE Solve the \gls{bs} precoder optimization under given $\boldsymbol{\theta}^{(i)}$, and obtain $\w_k^{(i+1)}$ \cite{Wei2024}.
    \STATE Solve problem $\mathcal{P}3$ under given $\w_k^{(i+1)}$ to obtain $\boldsymbol{\theta}^{(i + 1)}$
    \STATE $i = i + 1.$
\UNTIL $ \left| \tfrac{\min_{k} \, \mathrm{SINR}_{k}^{(i+1)} - \min_{k} \, \mathrm{SINR}_{k}^{(i)}}{\min_{k} \, \mathrm{SINR}_{k}^{(i+1)}} \right| \leq \epsilon $
\end{algorithmic}
\end{algorithm}
\end{footnotesize}

\textbf{Complexity Analysis:}
In our proposed algorithm presented in Alg. \ref{Alg1}, each iteration solves an \gls{sdr} for the precoder in $\mathcal{O}\bigl((KM)^{3.5}\bigr)$ and a feasibility problem via CVX in $\mathcal{O}(N)$ for the phase shifters. Repeating for $I_{\max}$ iterations gives a total complexity of $ \mathcal O\!\bigl(I_{\max}\bigl((KM)^{3.5} + N\bigr)\bigr)$.

\section{Simulation Results}
In this section, we provide the numerical results and analysis that show the effectiveness of our approach.

\subsection{Simulation parameters}
All components of the network, including \gls{bs}, \gls{lc}-\gls{ris} and \glspl{ue}, are positioned in the far-field relative to each other, with the \gls{lc}-\gls{ris} and \gls{bs} centered at $\p_{\mathrm{RIS}}$ and $\p_{\mathrm{BS}}$, respectively. In our simulations, we consider $1000$ vectors for Gaussian randomization for the \gls{sdr} in precoder optimization. The applied distance-dependent pathloss model is given by  $ P_\mathrm{L}(d_{\mathrm{t,r}}) = C_0 (d_0/d_{\mathrm{t,r}})^\sigma $, where $ C_0 = -61 \, \text{dB} $ at $ d_0 = 1 \, \text{m} $, $d_{\mathrm{t,r}}$ is the distance between transmitter and receiver, and $\sigma$ is the pathloss exponent \cite{Delbari2024temperature}. The remaining simulation parameters are summarized in Table~\ref{tab:system_parameters}. The performance of our \textit{proposed} scheme, presented in Alg.~\ref{Alg1}, is evaluated against three benchmark schemes:
\begin{enumerate*}[label=(\arabic*)]
  \item \textit{Temperature-neglecting}: In this benchmark, we design the phase shifts for the reference temperature while neglecting the impact of temperature variations;
  \item \textit{Random phase-shift}: Here, the \gls{lc}-\gls{ris} elements are assigned random phase shifts without optimization;
  \item \textit{Without RIS}: This scheme excludes any \gls{ris} from the system model and relies solely on the \gls{mrt} precoder at the \gls{bs}.
\end{enumerate*}
In addition to these benchmarks, we consider an upper bound that assumes access to the full range of phase shifts regardless of temperature constraint\footnote{While this Theoretical upper bound may not be practically achievable under temperature variations exceeding $T_\mathrm{r}$ in \gls{lc}-\gls{ris} systems, it nevertheless provides valuable insight into the performance gap between our proposed scheme and an ideal system.}.


\begin{table}[h!]
\caption{Simulation Parameters.}
\label{tab:system_parameters}
\centering
\resizebox{0.6\columnwidth}{!}{%
\begin{tabular}{c@{\hskip 10pt}c@{\hskip 10pt}c@{\hskip 10pt}c}
\toprule
\textbf{Parameter} & \textbf{Value} & \textbf{Parameter} & \textbf{Value} \\
\midrule
\text{Frequency} & $28$ GHz & $\sigma$ & $2$ \\
$P$ & $40$ dBm & $\epsilon$ & $10^{-3}$ \\
$\sigma_n^2$ & $-80$ dBm & $K_{\mathrm{d,k}}$  & $10$ dB \\
$M$ & $64$ & $K_{\mathrm{r}}$ & $10$ dB \\
$N$ & $40 \times 40$ & $K_{\mathrm{R,k}}$ & $10$ dB \\
$T_\mathrm{r}$ & $300$ K & $\p_{\mathrm{BS}} $ & $(0,20,4)$ \\
$T_c$ & $400$ K & $\p_{\mathrm{RIS}}$ & $ (0,0,4)$  \\
$T$ & $55^\circ\mathrm{C}$ & $K$ & $2$  \\
\bottomrule
\end{tabular}%
}
\end{table}

\subsection{Numerical results}
\begin{figure}[t]
\centering
\includegraphics[width=1\columnwidth,keepaspectratio]{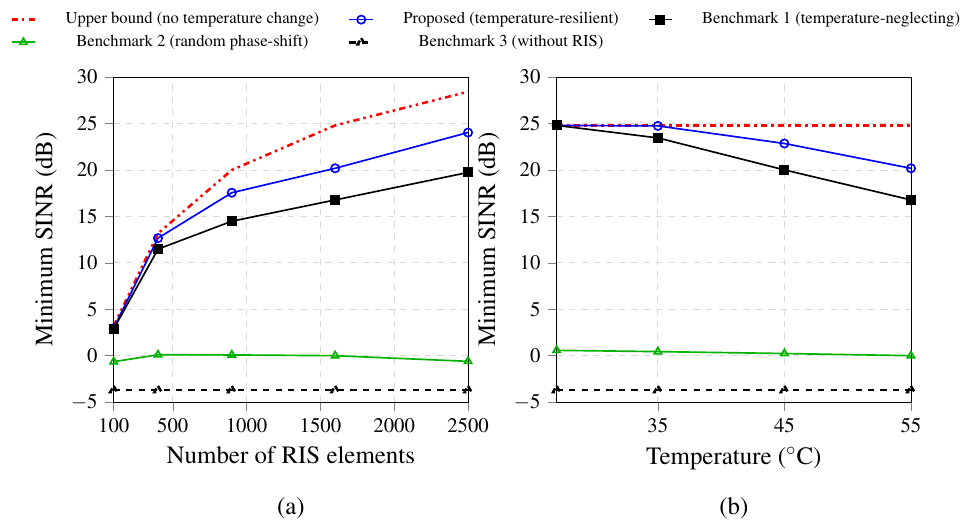}
\caption{ Minimum SINR performance comparison under different configurations. (a) Impact of the number of RIS elements on the minimum SINR. (b) Impact of ambient temperature (in $^\circ\mathrm{C}$) on the minimum SINR.}
\label{fig:merge}
\end{figure}

Fig.~\ref{fig:merge}a shows the minimum \gls{sinr} value across various \gls{lc}-\gls{ris} sizes. As expected, increasing the number of elements improves the minimum \gls{sinr} since more elements allow for more precise beamforming, which helps reduce interference and focus energy toward the \glspl{ue}. The \textit{proposed} scheme consistently outperforms the benchmark schemes across all \gls{lc}-\gls{ris} sizes, showing effective handling of temperature-induced phase-shift limitations. Moreover, although the \textit{benchmark $2$} experiences a significant drop in minimum \gls{sinr} performance, it still outperforms the \textit{benchmark $3$}, highlighting 
the importance of incorporating \gls{lc}-\gls{ris} in a multi-user downlink scenario. Furthermore, Fig.~\ref{fig:merge}b shows the impact of exceeding temperature $T_\mathrm{r}$ on the minimum \gls{sinr}. When $T > T_\mathrm{r}$, temperature fluctuations lead to a loss of full-range phase-shift functionality, causing the beams to disperse and thereby increasing the interference. \textit{Benchmark $1$}, which disregards these limitations, experiences a significant performance degradation, whereas the \textit{proposed} scheme effectively compensates for this limitation. 

\begin{figure}[t]
    \centering
    \includegraphics[width=1\columnwidth]{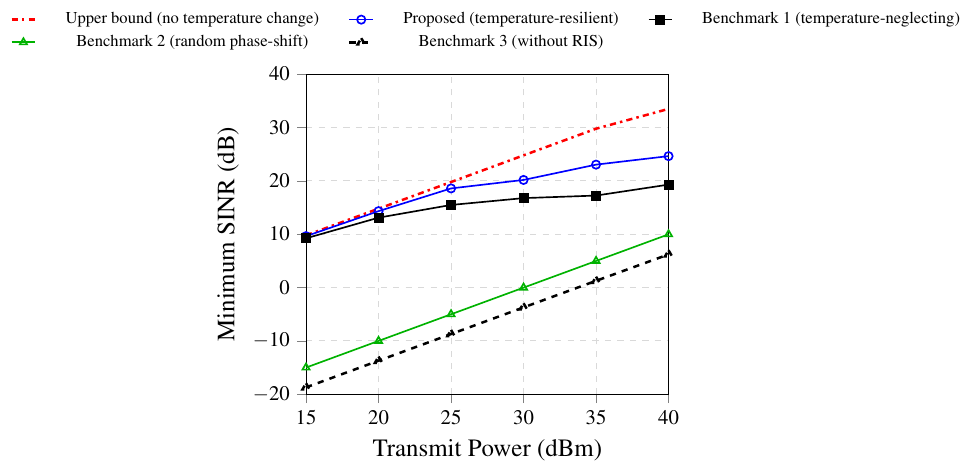}
    \caption{Impact of \gls{bs} transmit power (dBm) on the minimum SINR performance.}
    \label{fig:power}
\end{figure}

Lastly, Fig.~\ref{fig:power} illustrates minimum \gls{sinr} against varying \gls{bs} transmit power. The increase in minimum \gls{sinr} is due to the stronger signal being able to better overcome interference and noise as the transmit power is raised. However, this improvement may saturate depending on the system configuration, e.g., \cite[Fig.~6]{delbari2024nearfield}.
On the other hand, the minimum \gls{sinr} achieved by the \textit{proposed} scheme has superiority to all three benchmarks, indicating merely raising transmit power does not fully offset the degradation caused by ignoring temperature-induced phase-shift limitations (and it's also not energy efficient). Additionally, although the \textit{benchmark $2$} and \textit{benchmark $3$} schemes exhibit very poor performance overall, they experience significant improvements with increased transmit power; nonetheless, they still fall considerably short of the \textit{proposed} and \textit{benchmark $1$} schemes.
\vspace{-1pt}
\section{Conclusion}
In this paper, we studied the challenge of the phase shifter range shrinkage problem in \gls{lc}-\gls{ris}-assisted communication. We proposed a temperature-resilient phase-shift design using \gls{sdr} and \gls{sca} techniques based on mathematical modeling, demonstrating a significant improvement over the benchmark approaches. Our proposed method showed superior performance by optimizing phase shifts within temperature-constrained limits, ensuring an increased max-min \gls{sinr} level. Future work will involve exploring near-field effects for extremely large \gls{lc}-\gls{ris}, inhomogeneous temperature across \gls{lc}-\gls{ris} surface, and investigating multi-\gls{lc}-\gls{ris} configurations to enhance system performance further while coordinating temperature effects across multiple surfaces.

\begin{footnotesize}
\bibliography{ref}
\end{footnotesize}
\end{document}